\documentclass[english]{article}
\usepackage[T1]{fontenc}
\usepackage[latin9]{inputenc}
\usepackage{textcomp}
\usepackage{graphicx}

\makeatletter

\DeclareRobustCommand{\greektext}{%
  \fontencoding{LGR}\selectfont\def\encodingdefault{LGR}}
\DeclareRobustCommand{\textgreek}[1]{\leavevmode{\greektext #1}}
\DeclareFontEncoding{LGR}{}{}

\makeatother

\usepackage{babel}

\begin{document}

\title{Can the Podkletnov effect be explained by quantised inertia?}

\author{M.E. McCulloch%
\thanks{SMSE, University of Plymouth, PL4 8AA, U.K. mike.mcculloch@plymouth.ac.uk%
}}
\maketitle
\begin{abstract}
The Podkletnov effect is an unexplained loss of weight of between
0.05\% and 0.07\% detected in test masses suspended above supercooled
levitating superconducting discs exposed to AC magnetic fields. A
larger weight loss of up to 0.5\% was seen over a disc spun at 5000
rpm. The effect has so far been observed in only one laboratory. Here,
a new model for inertia that assumes that inertial mass is caused
by Unruh radiation which is subject to a Hubble-scale Casimir effect
(called MiHsC or quantised inertia) is applied to this anomaly. When
the disc is exposed to the AC magnetic field it vibrates (accelerates),
and MiHsC then predicts that the inertial mass of the nearby test
mass increases, so that to conserve momentum it must accelerate upwards
against freefall by 0.0029 m/s\textasciicircum{}2 or 0.03\% of g,
about half of the weight loss observed. With disc rotation, MiHsC
predicts an additional weight loss, but 28 times smaller than the
rotational effect observed. MiHsC suggests that the effect should
increase with disc radius and rotation rate, the AC magnetic field
strength (as observed), and also with increasing latitude and for
lighter discs.
\end{abstract}

\section{Introduction}

The Podkletnov effect, first observed by Podkletnov {[}1, 2{]} is
a small weight loss seen in test masses suspended above supercooled
levitating superconducting discs subjected to an AC magnetic field,
and spinning. The effect was independent of the masses\textquoteright{}
composition, and was not due to moving air since it persisted when
they were encased in glass. It was not magnetic since it persisted
when a metal screen was placed between the disc and the test masses.
Without disc rotation the effect produced a weight loss of between
0.05\% and 0.07\%, and with rotation, up to 0.5\%. The effect was
largest near the outer edge of the disc and greatest when the disc
was decelerated {[}2{]}. It was an apparent upwards force, and so
was different from the Tajmar effect {[}3{]} which is a rotational
acceleration seen close to supercooled rings, though in this paper
it is suggested that these two effects may be explained the same way.

Neither the Tajmar, nor the Podkletnov effect have been seen independently
in another laboratory. Some attempts have been made to reproduce the
latter {[}4, 5{]}, but these did not reproduce the experimental conditions
exactly. Given the potential significance of these experiments, an
exact replication is needed.

The author {[}6{]} proposed a model for inertia that could be called
a Modification of inertia resulting from a Hubble-scale Casimir effect
(MiHsC or Quantised Inertia). MiHsC assumes that the inertial mass
of an object is caused by Unruh radiation resulting from the acceleration
of surrounding matter, and that this radiation is subject to a Hubble-scale
Casimir effect: only Unruh waves that fit exactly into twice the Hubble
distance are allowed, which means that the Unruh waves are increasingly
disallowed as they get longer, leading to a gradual new loss of inertia
as acceleration reduces. In MiHsC the inertial mass ($m_{I}$) becomes

\begin{equation}
m_{I}=m\left(1-\frac{2c^{2}}{a\Theta}\right)\end{equation}

Where \textit{m} is the gravitational mass, \textit{c} is the speed
of light, \textit{\textgreek{J}} is the Hubble diameter and \textquoteleft{}\textit{a}\textquoteright{}
is the magnitude of the relative acceleration of surrounding matter.
MiHsC predicts the Pioneer anomaly {[}6{]} beyond 10 AU, the Earth
flyby anomalies quite well {[}7{]}, the Tajmar effect very well {[}8,
10{]}, cosmic acceleration {[}6, 9{]} and some aspects of galaxy rotation
{[}9{]}, all without adjustable parameters.

The Tajmar effect is of special interest here since it is an anomalous
acceleration observed by laser gyroscopes close to, but isolated from,
supercooled rings {[}3{]}. MiHsC predicts the Tajmar effect very well
{[}10{]}. When the ring accelerates the gyroscopes gain inertia by
MiHsC, and to conserve momentum with respect to the ring, they have
to move with the ring, just as observed. The Tajmar setup was similar
to that of Podkletnov (but involved no levitation) so in this paper
MiHsC is applied to the Podkletnov results {[}2{]}.

MiHsC (equation (1)) violates the equivalence principle (very slightly
for higher terrestrial accelerations), but not in a way that could
be detected by the usual torsion balance method. These experiments
measure the differential attraction of two balls on a cross bar suspended
on wire, towards distant masses (eg: the Sun) by detecting tiny twists
in the wire {[}11{]}. With MiHsC the two balls would have equal accelerations
with respect to the distant masses (being rigidly connected) so their
inertial masses should be modified equally by MiHsC, and there will
be no twist in the wire and no apparent violation of equivalence.

\section{Method}

We first consider a test mass (m) suspended over Podkletnov\textquoteright{}s
spinning disc and its conservation of momentum in the vertical direction

\begin{equation}
m_{2}v_{2}=m_{1}v_{1}\end{equation}

Where v is the velocity upwards. Replacing m using the quantised inertia
of {[}6{]} (equation (1)) gives

\begin{equation}
v_{2}\left(1-\frac{2c^{2}}{a_{2}\Theta}\right)=v_{1}\left(1-\frac{2c^{2}}{a_{1}\Theta}\right)\end{equation}

Rearranging

\begin{equation}
dv=\frac{2c^{2}}{\Theta}\left(\frac{v_{2}}{a_{2}}-\frac{v_{1}}{a_{1}}\right)\end{equation}

This is the change in velocity (\textit{dv}) of the test mass required
to conserve momentum following changes in the acceleration of its
surroundings from $a_{1}$ to $a_{2}$ which changes the inertia of
the test mass by MiHsC. A similar formula was derived by the author
{[}7{]} for the Earth flyby anomalies. Following {[}8{]} the $a_{1}$
represents all the initial accelerations (at time=1) near the test
mass. Each surrounding acceleration should be weighted by the mass
of the object accelerating divided by its distance squared {[}8, 10{]},
but this detail is not needed in this paper. Before the disc accelerates
there is little thermal acceleration because of the cryostat, and
the experiment was solidly fixed to the Earth\textquoteright{}s surface
so the mass of the planet itself causes no acceleration. However,
the Earth is spinning so the test mass and the fixed stars do mutually
accelerate, so $a_{1}=a_{s}$ where $a_{s}=v^{2}/r$, where r is the
distance from the spin axis ($r=r_{0}cos\varphi$, where $r_{0}$
is the Earth\textquoteright{}s radius: 6367500m and \textit{\textgreek{f}}
is the latitude at Tampere, Finland where the experiment took place:
$61.5^{o}N$) and \textit{v} is the Earth\textquoteleft{}s rotation
speed at $61.5^{o}N$ (v = 2\textgreek{p}r/86400 = 221m/s). Therefore
$a_{s}=0.016m/s^{2}$ plus a contribution from the acceleration due
to the Earth\textquoteright{}s orbit: $0.006m/s^{2}$ to give a total
acceleration of $0.022m/s^{2}$. Now $a_{2}$ is the environmental
acceleration when the disc is spinning which consists of a weighted
contribution from the fixed stars and the disc, but we can simplify
here since $a_{2}=v_{2}^{2}/r_{D}$ (where $r_{D}$ is the disc radius)
so the first term in equation (4) looks like $1/v_{2}$, so for high
rotational velocities we can ignore it. Therefore

\begin{equation}
dv=\frac{2c^{2}}{\Theta}\left(-\frac{v_{1}}{a_{s}}\right)\end{equation}

Since the inertial mass of the test mass has increased the \textit{dv}
acts to slow the original downwards velocity $v_{1}$ (the minus sign)
to conserve momentum (\textit{mv}). We can now differentiate equation
(5) with respect to time, assuming $a_{s}$ is constant, to get an
anomalous acceleration: \textit{da}. The $v_{1}$ becomes the acceleration
of the test mass with respect to the disc and this can be written
as \textquoteleft{}\textit{$-a_{d}$} \textquoteleft{} (the acceleration
of the disc with respect to the test mass).

\begin{equation}
da=\frac{2c^{2}}{\Theta}\left(\frac{a_{d}}{a_{s}}\right)\end{equation}

Therefore to conserve momentum the test mass must accelerate upwards
by \textit{da} to counter the increase in its inertial mass predicted
by MiHsC due to the disc\textquoteright{}s acceleration. The disc\textquoteright{}s
acceleration is rotational ($v^{2}/r$), where r is now the disc\textquoteright{}s
radius, and vibrational ($a_{v}$), the latter being caused by the
AC magnetic field so, if \textit{R} is the rotation rate in rpm

\begin{equation}
a_{d}=\frac{v^{2}}{r}+a_{v}=\frac{\left(R\times\frac{2\pi r}{60}\right)^{2}}{r}+a_{v}=\frac{4\pi^{2}R^{2}r}{3600}+a_{v}\end{equation}

Substituting equation (7) into equation (6) we get

\begin{equation}
da=\frac{2c^{2}}{\Theta}\left(\frac{4\pi^{2}R^{2}r}{3600a_{s}}+\frac{a_{v}}{a_{s}}\right)\end{equation}

In Podkletnov\textquoteright{}s experiment {[}2{]} first of all an
AC magnetic field was applied to the superconducting disc. The acceleration
($a_{v}$) on a superconductor of mass \textit{m} and area \textit{A}
from a magnetic field (\textit{B}, in Tesla) is

\begin{equation}
a_{v}=\frac{B^{2}A}{2\mu_{0}m}\end{equation}

Where $\mu_{0}$ is the permeability of free space ($4\pi\times10^{7}NA^{-2}$)
(The disc is being accelerated alternatively up and down, but with
MiHsC it is only the magnitude of the mutual acceleration that is
important, and not its direction, so the doubling of the vibrational
acceleration in the version of this paper to be published in Physics
Procedia (in press) is incorrect). To summarise, the predicted MiHsC
acceleration is

\begin{equation}
da=\frac{2c^{2}}{\Theta}\left(\frac{4\pi^{2}R^{2}r}{3600a_{s}}+\frac{B^{2}A}{2\mu_{0}ma_{s}}\right)\end{equation}

\section{Results}

In this section the predictions of MiHsC (equation (10)) are compared
with the observations of Podkletnov {[}2{]} for cases without and
with disc rotation.

The disc was initially subject to an AC magnetic field of 2 Tesla,
and was vibrating, but not rotating. So substituting numbers into
equation (10) (the mass of the disc was 0.95 kg, {[}12{]} and its
radius was 0.135m):

\begin{equation}
da=6.7\times10^{-10}\times\left(0+\frac{2^{2}\times\left(\pi\times0.135^{2}\right)}{2\times4\pi\times10^{-7}\times0.95\times0.022}\right)=0.0029\pm0.00025m/s^{2}\end{equation}

This \textit{da} is the anomalous acceleration (weight loss) predicted
by MiHsC. It is 0.03\% (\textpm{}0.0025\%) of the acceleration due
to gravity ($g=9.8m/s^{2}$) and this is about half of the observed
weight loss {[}2{]} which was between 0.05\% and 0.07\% of \textit{g}.
The error bars on the prediction were calculated assuming a 9\% error
due to uncertainties in the Hubble constant (and therefore \textit{\textgreek{J}}
in equation (10)).

Adding now the rotation of 5000 rpm which was applied to the disc
by Podkletnov {[}2{]}, this increases the acceleration of the nearby
disc and therefore, by MiHsC, the inertial mass of the test mass.
The predicted weight loss is now

\begin{equation}
da=6.7\times10^{-10}\times\left(\frac{4\pi^{2}\times5000^{2}\times0.1375}{3600\times0.022}+\frac{191842.1}{0.022}\right)=0.0041m/s^{2}\end{equation}

This is 0.042\% of the acceleration due to gravity (weight) and only
a small increase from the result with no rotation. For a rotating
disc, Podkletnov {[}2{]} observed a weight loss of between 0.3\% and
0.5\% of the weight. So MiHsC underestimates the observed increase
of the anomaly due to ring rotation by about a factor of 28.

\section{Discussion}

Figure 1 shows a much simplified view of the Podkletnov setup from
the point of view of the North Pole. The disc is shown within its
cryostat. The test mass (m) hangs above it and feels an acceleration
or weight of g downwards towards the disc. It may be useful to picture
the test mass falling towards the disc with an acceleration \textit{g}.
When the disc\textquoteright{}s environment is cooled, nearby accelerations
are reduced so Unruh waves become longer, a greater proportion are
disallowed by the Hubble-scale Casimir effect of MiHsC and the inertial
mass of the test mass slightly decreases, so its weight should appear
to increase as the Earth\textquoteright{}s pull will have more effect
(no data is available from the cooling process to test this). When
the AC field vibrates the disc this adds high accelerations to the
system again, the Unruh waves shorten, fewer are disallowed by MiHsC
so the inertial mass of the test mass increases and to conserve momentum
the test mass has to accelerate up (to counter the downwards acceleration,
\textit{g}) by 0.03\% of \textit{g} (the vector da on Figure 1), so
the test mass seems to lose 0.03\% of its weight. A free-falling test
mass would see its inertia increase, so its acceleration downwards
due to gravity would decrease. The assumption here is that the same
weight loss occurs for the suspended (static) test mass.

In summary, instead of the \textquoteleft{}gravity shielding\textquoteright{}
proposed by Podkletnov {[}2{]} the process proposed here is an increase
in inertial mass due to MiHsC, which makes the test mass less responsive
to the downward acceleration of gravity: an apparent loss of weight.
The prediction by MiHsC of the effect with only the AC magnetic field
is about half of the observed weight loss. However, the prediction
of the extra effect due to spin is 28 times smaller than observed.
It could be that the vibration and rotation couple in some way to
boost the acceleration and thereby the effect of MiHsC. According
to MiHsC the anomalous effect should vanish outside the cryostat,
which disagrees with the extended vertical column of the weight loss
observed by {[}2{]}.

The author {[}10{]} applied MiHsC to the Tajmar effect using a conservation
of momentum relative to the ring. In that case as the ring rotated
(accelerated) and the gyro\textquoteright{}s inertial mass increased,
their velocity relative to the ring had to decrease, so the gyros
had to move with the ring. In this case the sudden acceleration of
the disc increases the inertial mass of the hanging test mass and
to conserve momentum it has to accelerate upwards against freefall.
The consistency of these results need to be checked further.

If MiHsC is the correct explanation, then it should be possible to
produce the Podkletnov effect without a superconductor, and, referring
back to equation (10), there are several ways to enhance it. First
of all the rotation rate (\textit{R}) could be increased, however
there would be a limitation since the inertial mass can approach \textit{m}
(the gravitational mass) but cannot go any higher (without somehow
the production of extra synthetic Unruh radiation). Podkletnov {[}2{]}
observed that the anomalous effect increased with the rotation rate.
What is not included in the above equation is the acceleration caused
by a change in the rotation rate of the disc, which should further
increase the inertial mass of the test mass by MiHsC and enhance the
effect. An effect like this, upon deceleration, was hinted at by Podkletnov.

The radius of the disc (\textit{r} in the formula, which also affects
\textit{A}) could be increased. Indeed Podkletnov {[}2{]} noticed
that the anomalous effect was largest at the outer edge of the disc,
as suggested by the first term of equation (10). One problem with
increasing the size of the disc is that the fabrication of large discs
is difficult.

The AC magnetic field strength (\textit{B}) could be increased. This
would be an effective way to increase the effect since \textit{B}
is a squared parameter in equation (10), but this may induce damaging
vibrations.

One interesting possibility is that the value of \textquoteleft{}$a_{s}$\textquoteright{}
(the acceleration with respect to the fixed stars) could be reduced
by moving the experiment to a higher latitude or to a frame moving
with the fixed stars, so when extra acceleration is added by vibration
or spinning the disc the inertial mass gain, and \textit{da}, are
larger. Finally, the mass of the disc \textquoteleft{}\textit{m}\textquoteright{}
in equation (10) could be reduced to increase the vibrational effect.

\section{Conclusions}

A new model for inertia (MiHsC) that assumes that inertia is due to
Unruh radiation which is subject to a Hubble-scale Casimir effect
was applied to the anomalous weight loss (Podkletnov effect) observed
over supercooled superconducting discs which are: Case 1) subject
to an AC magnetic field and Case 2) also spinning at 5000rpm.

For case 1, MiHsC predicts a loss of weight of 0.03\%, about half
of the observed weight loss which was between 0.05\% and 0.07\%. However,
for case 2 the predicted weight loss was 0.042\%, an increase which
is 28 times smaller than the extra weight loss observed.

MiHsC predicts that the Podkletnov effect should increase with disc
radius and speed, AC magnetic field strength, latitude (or for a system
rotating with the fixed stars), and also increase for lighter discs.

\section*{Acknowledgements}

Thanks to G.A. Robertson for organising and editing the SPESIF-2011
proceedings, an anonymous reviewer for useful comments, and to J.
Wharton and B. Kim for encouragement.

\section*{References}

{[}1{]}. Podkletnov, E. E. and Nieminen, R., A possibility of gravitational
shielding by bulk YBa2Cu3O7-x superconductor, \textit{Physica C} 1992
203:441-444.

{[}2{]}. Podkletnov, E. E., Weak gravitational shielding properties
of composite bulk YBa2Cu3O7-xsuperconductor below 70K under e.m. field,
\textit{Arxiv}:cond-mat/9701074v3 1997.

{[}3{]}. Tajmar, M., Plesescu, F. and Seifert, B., Anomalous fiber
optic gyroscope signals observed above spinning rings at low temperature,
\textit{J. Phys. Conf. Ser.} 2009 150:032101.

{[}4{]}. Li, N., Noever, D., Robertson, T., Koczor, R. and Brantley,
W., Static Test for a Gravitational Force Coupled to Type II YBCO
Superconductors, \textit{Physica C} 1997 281:260-267.

{[}5{]}. Woods, C., Cooke, S., Helme, J. and Caldwell, C., Gravity
Modification by High Temperature Superconductors. Joint Propulsion
Conference, AIAA 2001-3363 2001.

{[}6{]}. McCulloch, M. E., Modelling the Pioneer anomaly as modified
inertia, \textit{MNRAS} 2007 376(1):338-342.

{[}7{]}. McCulloch, M. E., Modelling the flyby anomalies using a modification
of inertia, \textit{MNRAS-letters} 2008 389(1):L57-60.

{[}8{]}. McCulloch, M. E., Can the Tajmar effect be explained using
a modification of inertia? \textit{EPL} 2010 89:19001.

{[}9{]}. McCulloch, M. E., Minimum accelerations from quantised inertia,
\textit{EPL} 2010 90:29001.

{[}10{]}. McCulloch, M. E., The Tajmar effect from quantised inertia,
\textit{EPL} 2011 95:39002 (http://arxiv.org/abs/1106.3266).

{[}11{]}. Gundlach, J. H., Schlamminger, S., Spitzer, C. D. and Choi,
K.-Y., Laboratory test of Newton's second law for small accelerations,
\textit{PRL} 2007 98:150801.

{[}12{]}. Solomon, B., Reverse engineering Podkletnov\textquoteright{}s
experiments, in these proceedings of the Space, Propulsion \& Energy
Sciences International Forum (SPESIF-2011), edited by G. A. Robertson,
\textit{Physics Procedia}, Elsevier Press. 2011.

\section*{Figures}

\noindent \begin{center}
\includegraphics[scale=1.2]{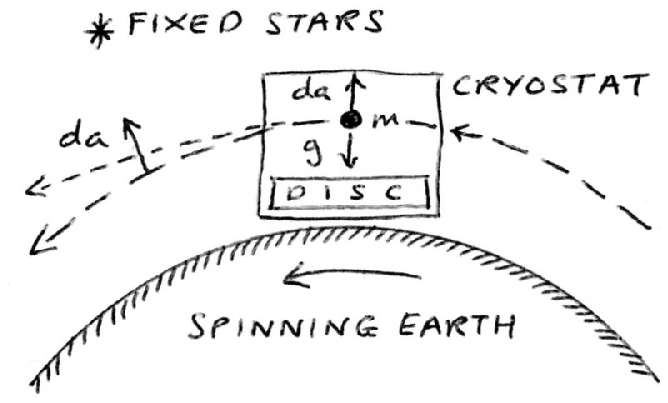}
\par\end{center}

\noindent Figure 1. A schematic view from the North Pole showing the
cryostat, the disc and the test mass (\textit{m}). They rotate eastward
(left) with the Earth along the long-dashed line. When the test mass
gains inertia by MiHsC it accelerates upwards (\textit{da}). Free
objects (air) would rise following the short-dashed line.
\end{document}